\documentclass{IEEEtran}
\usepackage{framed,multirow}
\usepackage{cite}
\usepackage{amsmath,amssymb,amsfonts}
\usepackage{algorithmic}
\usepackage{graphicx}
\usepackage{textcomp}
\usepackage{multirow}
\usepackage{comment}
\usepackage{tabu}
\usepackage{makecell}
\usepackage[colorlinks=true]{hyperref}
\usepackage{siunitx}

\begin{document}
\title{Multi-Modal MRI Reconstruction Assisted with Spatial Alignment Network}

\author{Kai Xuan, %
Lei Xiang, %
Xiaoqian Huang, %
Lichi Zhang, %
Shu Liao, %
Dinggang Shen*, %
and Qian Wang*
\thanks{Kai Xuan, Lei Xiang, and Lichi Zhang
are with the School of Biomedical Engineering,
Shanghai Jiao Tong University, Shanghai, China.
(e-mail: \{kaixuan, xianglei\_15, lichizhang\}@sjtu.edu.cn)}
\thanks{Xiaoqian Huang and Shu Liao
are with Shanghai United Imaging Intelligence Co., Ltd., Shanghai, China.
(e-mail: \{xiaoqian.huang, shu.liao\}@united-imaging.com)}
\thanks{Dinggang Shen is with the School of Biomedical Engineering,
ShanghaiTech University, Shanghai, China.
He is also with Shanghai United Imaging Intelligence Co., Ltd.,
Shanghai, China.
(e-mail: dinggang.shen@gmail.com)}
\thanks{Qian Wang is with the School of Biomedical Engineering,
ShanghaiTech University, Shanghai, China.
(e-mail: wangqian2@shanghaitech.edu.cn)}
\thanks{This research was supported by the grants from
the National Key R\&D Program of China (2018YFC0116400),
National Natural Science Foundation of China (62131015),
Science and Technology Commission of Shanghai Municipality
(19QC1400600 and 21010502600),
and the Key R\&D Program of Guangdong Province, China (2021B0101420006).}
}

\maketitle

\begin{abstract}
In clinical practice, multi-modal magnetic resonance imaging (MRI)
with different contrasts is usually acquired in a single study
to assess different properties of the same region of interest in the human body.
The whole acquisition process can be accelerated
by having one or more modalities under-sampled in the $k$-space.
Recent research has shown that,
considering the redundancy between different modalities,
a target MRI modality under-sampled in the $k$-space
can be more efficiently reconstructed
with a fully-sampled reference MRI modality.
However, we find that the performance of 
the aforementioned multi-modal reconstruction can be negatively affected
by subtle spatial misalignment between different modalities,
which is actually common in clinical practice.
In this paper, we improve the quality of multi-modal reconstruction
by compensating for such spatial misalignment with a spatial alignment network.
First, our spatial alignment network estimates the displacement
between the fully-sampled reference and the under-sampled target images,
and warps the reference image accordingly.
Then, the aligned fully-sampled reference image joins
the multi-modal reconstruction of the under-sampled target image.
Also, considering the contrast difference between the target and reference images,
we have designed a cross-modality-synthesis-based registration loss
in combination with the reconstruction loss,
to jointly train the spatial alignment network and the reconstruction network.
The experiments on both clinical MRI and multi-coil $k$-space raw data demonstrate the superiority and robustness of the multi-modal MRI reconstruction empowered with our spatial alignment network.
Our code is publicly available at
\url{https://github.com/woxuankai/SpatialAlignmentNetwork}.
\end{abstract}


\section{Introduction}
Magnetic resonance imaging (MRI) is a non-invasive
and radiation-free diagnostic technology that has been widely used in
clinical practice.
However, limited by the device and imaging protocols,
MR scan is relatively slow,
and the acceleration of MRI acquisition has been an
important and everlasting research topic since its invention.
A feasible remedy is to exploit the redundancy between signals
received from different coils in parallel imaging~\cite{
pruessmann_sense_1999,griswold_generalized_2002}.
Also, compressed sensing MRI (CS-MRI)~\cite{lustig_sparse_2007}
allows for accurate reconstruction from the signals that are highly
sparse in the $k$-space with respect to the Nyquist-Shannon sampling
criterion.

Aside from accelerating MRI acquisition by exploiting the intrinsic
redundancy within a single MR image, the common information coupled
and shared by different MRI sequences draws a lot of attention. In
clinical practice, MR images of different contrasts are usually acquired
in the same study to reflect different properties of the same region
of interest (ROI) and to facilitate precise diagnosis.
Recent research~\cite{xiang_ultra-fast_2018, kim_improving_2018}
has shown that multi-modal MRI reconstruction
is better at reconstructing an under-sampled MR sequence
(i.e., the \textit{target modality})
with auxiliary information from another fully-sampled
\textit{reference modality} in the same study.
With the target acquisition under-sampled
in the $k$-space, the overall time cost of the whole study in MR scanning
can thus be reduced.

\begin{figure*}[!t]
\centerline{\includegraphics[width=\textwidth]{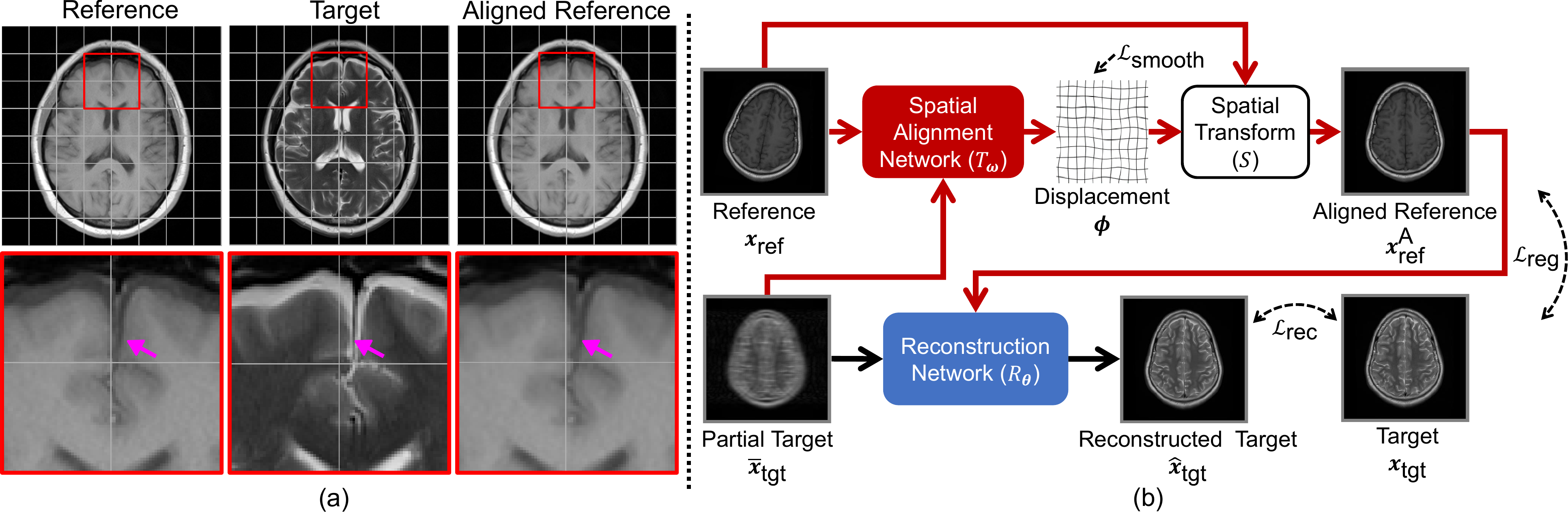}}
\caption{A real case demonstrating the existence of spatial misalignment (a),
and the overview of the proposed method (b).
In (a), a real case of multi-modal
MRI acquired for the diagnostic purpose demonstrates the existence
of spatial misalignment (highlighted by arrows) between the reference
(T1-weighted) and the target (T2-weighted) images.
The aligned reference image is also available to show
the effect of our proposed spatial alignment network.
In (b), a spatial alignment network is integrated
into the multi-modal MRI reconstruction pipeline
to compensate for the spatial misalignment between the fully-sampled
reference image and the under-sampled target. The data flow for the
conventional deep-learning-based reconstruction is shown in black
arrows, and the red arrows are for additional data flow related to
our proposed spatial alignment network.}
\label{fig_overview}
\end{figure*}

However, we find that the spatial misalignment between modalities,
which is subtle and previously ignored,
can non-negligibly weaken the final reconstruction quality
of the target modality.
Such spatial
misalignment is prevalent between individual sequences of the same
study. A real example can be found in Fig.~\ref{fig_overview}(a),
where the consecutively acquired T1-weighted and T2-weighted images
from the same subject are shown. One may easily notice the misalignment
of the corresponding anatomic structures as highlighted by the arrows.

The reason behind such misalignment can be complex. Although the subjects
are typically instructed to keep still during MRI acquisition, a person
even without any MRI knowledge can easily notice the gap between two
sequences, e.g., due to pause and altered noises of the scanner.
Thus, one may then tend to relax and have subtle movement,
which can lead to inconspicuous yet mostly inevitable motion
between the sequences,
even though external stabilization measures are usually deployed.

However, the spatial
misalignment between MRI modalities of the same study is mostly ignored in the literature.
With CS-MRI, Lai \textit{et al.}~\cite{lai_sparse_2017}
utilized multi-modal image registration to align the reference modality with the intermediate reconstruction result of the under-sampled target modality,
while the conventional image registration
may not be suitable for the under-sampled MRI.
In the recent deep
learning era, many reported experiments of multi-modal MRI reconstruction
are conducted on carefully registered multi-contrast MRI~\cite{
xiang_deep-learning-based_2019,
zhou_dudornet_2020, lyu_multi-contrast_2020, dar_prior-guided_2020}.
Although image pre-processing can effectively suppress the misalignment
for training and validation in labs, such issues can hardly be avoided
in real scenarios.

Enlarging model capacity with a deeper neural network
is a universal solution towards better performance,
but it lacks insight into the spatial misalignment issue
and suffers from  diminishing marginal utility
in multi-modal reconstruction~\cite{sun_deep_2019}.
Alternatively, in this work, we propose an effective solution to explicitly
mitigate the negative impact of the spatial misalignment. Particularly,
we use a plug-in spatial alignment network to align the reference
MRI with the zero-filled target image, in combination with the multi-modal
MRI reconstruction network. An overview of our proposed method is
shown in Fig.~\ref{fig_overview}(b). First, the spatial alignment
network estimates the motion between the fully-sampled
reference MRI and the under-sampled target. Then, the warped reference
image, instead of the misaligned original one, is concatenated with
the under-sampled target and fed into the reconstruction network,
which is then able to yield the estimated high-quality target image.

Our contributions are summarized as follows:
\begin{itemize}
\item To improve multi-modal MRI reconstruction,
we propose to use the spatial alignment network
that explicitly compensates for the subtle spatial misalignment
between the reference and the target modalities.
\item To optimize the spatial alignment network,
we propose to combine the cross-modality-synthesis-based loss
with the image reconstruction loss,
which leads to improved registration and reconstruction quality.
\item To demonstrate the superiority and robustness of the proposed method,
we propose to conduct extensive experiments
on both real-valued clinical MRI and complex-valued multi-coil data.
\end{itemize}

\section{Related Works}\label{section_related_works}
In this section, we introduce the development of MRI reconstruction
in Section~\ref{section_related_recon} firstly,
and then the idea of improving target image reconstruction quality
with auxiliary information from the reference image
in Section~\ref{section_related_ref}.
Finally, we discuss multi-modal medical image alignment
in Section~\ref{section_related_alignment}.

\subsection{MRI Reconstruction}
\label{section_related_recon} To recover fully-sampled MR images
from partially acquired $k$-space signals, image reconstruction plays
a key role especially concerning the final image quality. Basic solutions
including linear filtering or zero-filling are used to recover the
missing $k$-space signals, while those relatively simple methods
tend to bring in artifacts in the reconstructed MR images.
Compressed sensing (CS)~\cite{donoho_compressed_2006}
and its application to MRI~\cite{lustig_sparse_2007}
are the milestone of under-sampled MRI reconstruction.
The technique has effectively accelerated MRI acquisition.

Deep-learning-based methods emerge as state-of-the-art in MRI reconstruction.
In the pilot work of Wang \textit{et al.}~\cite{wang_accelerating_2016},
a convolutional neural network (CNN) is trained offline to model the
mapping from zero-filled under-sampled MR images to the corresponding
fully-sampled ones. Later, the techniques such as data consistency
layers~\cite{schlemper_deep_2018} and recurrent neural networks
(RNNs)~\cite{qin_convolutional_2019} are used to further improve
the reconstruction quality. While most works follow the image-space modeling
paradigm used in CS-MRI, Han \textit{et al.}~\cite{han_k_2020}
proposed to estimate the missing MR signals in the $k$-space where
MRI is physically acquired.
Further, Zhang \textit{et al.}~\cite{zhang_dual-domain_2019}
processed MRI signals in both image space and $k$-space.
Another interesting
deep-learning-based solution for MRI reconstruction is to directly
estimate the fully-sampled MR images from partially acquired $k$-space
signals without invoking Fourier transform explicitly~\cite{zhu_image_2018}.
Recently, the NYU fastMRI challenges,
especially the large-scale public datasets,
have greatly promoted the development of reconstruction algorithms~\cite{%
zbontar_fastmri_2019,knoll_fastmri_2020,
knoll_advancing_2020,muckley_results_2021}.

While the works mentioned above focus on the reconstruction of
a single MRI modality independently,
our proposed method takes multiple MRI modalities into consideration
during the multi-modal reconstruction.

\subsection{Reference-based MRI Reconstruction}
\label{section_related_ref}
With physiology-related redundancy between
MR images, reconstructing under-sampled MRI with auxiliary information
from reference images is promising. Regarding multiple visits
of the same patient, Souza \textit{et al.}~\cite{souza_enhanced_2020}
proposed to accelerate MRI follow-ups with early intra-subject images.
Weizman \textit{et al.}~\cite{
weizman_compressed_2015, weizman_reference-based_2016}
proposed a more flexible solution for the case when
the reference was not fully available.
Schlemper \textit{et al.}~\cite{schlemper_deep_2018}
took advantage of the redundancy between adjacent temporal frames
of dynamic MRI by learning the temporal correlation
with 3D convolution (2D + time).
Moreover, the redundancy between
spatially neighboring 2D slices in a 3D volume was used in Hirabayashi
\textit{et al.}~\cite{hirabayashi_compressed_2015}, to improve
the reconstruction quality by fully-sampling selected slices
while under-sampling the rest. 

The reconstruction of a certain MRI modality can be enhanced with
the help from other modalities, as a typical MRI study consists of
multiple sequences and they are usually acquired sequentially.
Majumdar \textit{et al.}~\cite{majumdar_joint_2011}
and Huang \textit{et al.}~\cite{huang_fast_2014}
jointly reconstructed T1- and T2-weighted images of the same ROI
from under-sampled MRI signals.
For learning-based methods, Song \textit{et al.}~\cite{song_coupled_2020}
proposed coupled dictionary learning for multi-contrast MRI reconstruction.
More recently, with deep neural networks, Kim \textit{et al.}~\cite{
kim_improving_2018}
and Xiang \textit{et al.}~\cite{xiang_ultra-fast_2018} proposed
to speed up T2-weighted acquisition by integrating T1-weighted
image as the reference into the reconstruction process.
More techniques,
such as progressive neural networks~\cite{lyu_multi-contrast_2020},
generative adversarial networks~\cite{dar_prior-guided_2020},
and dilated convolutions~\cite{zhou_dudornet_2020},
are also investigated to improve multi-modal MRI reconstruction.
Also, note that multi-modal reconstruction is not only limited to MRI.
The idea has been successfully used in various tasks,
e.g., enhancing low-dose positron emission tomography (PET) quality
with corresponding MRI
to reduce the radiation risk of PET/MRI scanning~\cite{xiang_deep_2017}.

Different from merging information from varying references,
our proposed method focuses on mitigating the spatial misalignment between the target and reference images,
which is mostly ignored by literature reports
but can hardly be avoided in real scenarios.

\subsection{Multi-Modal MRI Alignment}
\label{section_related_alignment}
Image registration is a classical topic in medical image analysis,
and it is developing fast nowadays when enabled with deep learning.
Traditional methods take registration as an optimization problem,
and the displacement field is optimized iteratively with an objective
function consisting of the image similarity loss and the regularization term.
Such methods are popular, including ANTs~\cite{avants_symmetric_2008},
Demons~\cite{thirion_image_1998,vercauteren_diffeomorphic_2009},
and SPM~\cite{hellier_inter-subject_2002}.
In the alternative, learning-based methods,
especially for deep-learning-based registration,
the complex displacement field can be directly learned
between the fixed and the moving images.
Early works train deep neural networks to predict the displacement fields
in a supervised manner~\cite{yang_quicksilver_2017}.
And later, Balakrishnan~\textit{et al.}~\cite{balakrishnan_unsupervised_2018}
proposed to train registration networks in an unsupervised manner.
While image registration methods usually focus on fully-sampled images,
Kustner~\textit{et al.}~\cite{kustner_lapnet_2021} showed the feasibility
to directly register under-sampled MR images.
They also explored the way to estimate the displacement fields
in Fourier space instead of image space.

It is critical to design accurate and robust metrics
to quantify image similarity,
especially in the case of multi-modal or multi-contrast registration.
However, such metrics are challenging to design in computer vision
due to the huge appearance gap across different image modalities.
Mutual information (MI), describing the pixel- or voxel-wise
statistical dependency between corresponding intensities of the images
to be aligned~\cite{thevenaz_optimization_2000},
has been successfully applied to various multi-modal image registration
tasks, e.g., PET-computed tomography (CT)~\cite{mattes_pet-ct_2003},
and PET-MRI registration~\cite{smith_advances_2004}. Later, feature-based
similarity metrics such as modality independent neighborhood descriptor
(MIND)~\cite{heinrich_mind_2012} take use of high-order appearance
features to establish spatial correspondences. With deep learning,
Fan \textit{et al.}~\cite{fan_adversarial_2019} proposed to measure
the image similarity with the generative adversarial network (GAN),
but paired images were required in their work as training samples.
Cao \textit{et al.}~\cite{cao_region_2018} circumvented
the direct similarity measurement with cross-modality synthesis. They
thus could replace the difficult calculation of multi-modal similarity
with a much simpler mono-modal metric. Further, to prevent possible
spatial distortion introduced by cross-modality synthesis,
the geometric-preserving technique was adopted
in Arar \textit{et al.}~\cite{arar_unsupervised_2020}.

Similar to Arar \textit{et al.}~\cite{arar_unsupervised_2020},
our proposed method attains
geometric-preserving cross-modality-synthesis-based registration
through the spatial alignment network.
Note that our registration handles under-sampled MR images,
while existing registration methods usually
focus on fully-sampled MRI.

\section{Methods}
\label{section_methods} The framework of our proposed method is illustrated
in Fig.~\ref{fig_overview}(b). It includes (1) the spatial alignment
network to register the reference image with the under-sampled
target, and (2) the reconstruction network to restore the target modality
image of high quality. Three major loss functions play
critical roles in our framework. Other than the commonly used smoothness
constraint $\mathcal{L}_{\mathrm{smooth}}$ upon the displacement field,
we also calculate the reconstruction loss $\mathcal{L}_{\mathrm{rec}}$
and the registration loss $\mathcal{L}_{\mathrm{reg}}$ to train the
spatial alignment network and the reconstruction network jointly.

The rest of Section~\ref{section_methods} will be arranged as follows.
Section~\ref{section_methods_mm} presents the mathematical formulation
of multi-modal MRI reconstruction aided by the spatial alignment network.
Then, the three loss functions,
i.e., the smoothness constraint $\mathcal{L}_{\mathrm{smooth}}$,
the reconstruction loss $\mathcal{L}_{\mathrm{rec}}$ and the registration
loss $\mathcal{L}_{\mathrm{reg}}$,
are described in Section~\ref{section_methods_opt}.
Finally, in Section~\ref{section_methods_details}, we introduce
more implementation details.

\subsection{Multi-Modal MRI Reconstruction}
\label{section_methods_mm} In this subsection,
we first formulate the learning-based
MRI reconstruction, and then introduce the multi-modal MRI reconstruction.
Next, we discuss how to integrate the proposed spatial alignment network
into the multi-modal MRI reconstruction pipeline.

\subsubsection{MRI Reconstruction}
The MRI acquisition can be perceived as sampling in a full $k$-space.
Taking 2D MRI of the image shape of $N_{x}\times N_{y}$ for example,
with a flattened complex-valued fully-sampled MRI image
$\boldsymbol{x}\in\mathbb{C}^{N}$
where $N=N_{x}\times N_{y}$, the fast MRI acquisition process can
be expressed as $\boldsymbol{y}=\boldsymbol{F}_{u}\boldsymbol{x}$,
where $\boldsymbol{y}\in\mathbb{C}^{M}$ is the under-sampled
MR signals in the $k$-space with the sampling ratio $\frac{M}{N}$,
($M\leq N$).
$\boldsymbol{F}_{u}\in\mathbb{C}^{M\times N}$,
being consistent with notations in Lustig
\textit{et al.}~\cite{lustig_sparse_2007},
is a matrix representing the whole MRI under-sampling process which usually
consists of the Fourier transform and the $k$-space under-sampling.
Note that the number of coils is ignored here for simplicity, though the
above formulation can be easily extended to the multi-coil setting.

Without loss of generality, the deep-learning-based MRI reconstruction,
which estimates the fully-sampled MRI $\boldsymbol{x}$
from the under-sampled $k$-space signals $\boldsymbol{y}$,
can be denoted as $R_{\boldsymbol{\theta}}$.
Here, for a modern MRI reconstruction method like
the End-to-End Variational Networks (E2E-VarNet)~\cite{sriram_end-to-end_2020},
$R_{\boldsymbol{\theta}}$ usually contains a serial of neural network blocks
and data consistency layers~\cite{schlemper_deep_2018},
and $\boldsymbol{\theta}$ is the union set of all learnable parameters.
Before feeding to $R_{\boldsymbol{\theta}}$,
usually the raw acquisition $\boldsymbol{y}$ is pre-processed with
$\boldsymbol{F}_{\mathrm{0-fill}}^{-1}\in\mathbb{C}^{N\times M}$,
which represents zero-filling of the missing MR signals and inverse
Fourier transform mapping the $k$-space data back to the image space.
In this way, one can have the under-sampled MR image
$\bar{\boldsymbol{x}}=\boldsymbol{F}_{\mathrm{0-fill}}^{-1}\boldsymbol{y}$,
where $\bar{\boldsymbol{x}}\in\mathbb{C}^{N}$.
Next, as aliasing ghosts typically exist in $\bar{\boldsymbol{x}}$,
the reconstruction model $R_{\boldsymbol{\theta}}$
should be capable of restoring the fully-sampled MRI
from the under-sampled one, i.e.,
$\hat{\boldsymbol{x}}=R_{\boldsymbol{\theta}}\left(\bar{\boldsymbol{x}}\right)$.
As a result, $R_{\boldsymbol{\theta}}$ is often referred as
the de-aliasing  model.

\subsubsection{Multi-Modal MRI Reconstruction}
The target MR image $\boldsymbol{x}_{\mathrm{tgt}}$ can be better
reconstructed with the help of the reference, such as an image of
the same ROI yet of different contrast acquired in
the same study. Extracting common information from the reference image
of a different contrast is non-trivial. And in the context of deep learning,
this process is expected to be completed in a sophisticated data-driven
and encoding-decoding way.

In early works~\cite{xiang_ultra-fast_2018,kim_improving_2018},
the fully-sampled reference modality
$\boldsymbol{x}_{\mathrm{ref}}^{\mathrm{A}}$
and the under-sampled target modality $\bar{\boldsymbol{x}}_{\mathrm{tgt}}$
are concatenated in the channel dimension,
and fed into a single reconstruction model $R_{\boldsymbol{\theta}}$.
It is expected that $\boldsymbol{x}_{\mathrm{ref}}^{\mathrm{A}}$
can help reconstruct $\boldsymbol{x}_{\mathrm{tgt}}$, as the
network should fuse individual channels of inputs and feature maps
through forward convolution. As a summary, the multi-modal MRI reconstruction
process can be formulated as
$\hat{\boldsymbol{x}}_{\mathrm{tgt}}
=R_{\boldsymbol{\theta}}\left(\bar{\boldsymbol{x}}_{\mathrm{tgt}},\boldsymbol{x}_{\mathrm{ref}}^{\mathrm{A}}\right)$.
The superscript $\mathrm{A}$ of $\boldsymbol{x}_{\mathrm{ref}}^{\mathrm{A}}$
indicates that the reference image is well-aligned to the target modality
virtually.

\subsubsection{Spatial Alignment Network}
In this work, we propose to integrate a spatial alignment network
$T_{\boldsymbol{\omega}}$ into the multi-modal MRI reconstruction, in
order to estimate and compensate for the subtle spatial misalignment between
the target and the reference modalities. First, the spatial alignment
network explicitly estimates the displacement field $\boldsymbol{\phi}$
between the under-sampled target modality $\bar{\boldsymbol{x}}_{\mathrm{tgt}}$
and the fully-sampled reference modality $\boldsymbol{x}_{\mathrm{ref}}$
following 
\begin{equation}
\boldsymbol{\phi}=T_{\boldsymbol{\omega}}\left(\bar{\boldsymbol{x}}_{\mathrm{tgt}},\boldsymbol{x}_{\mathrm{ref}}\right),
\end{equation}
where $\boldsymbol{\phi}\in\mathbb{R}^{2\times N}$ in our implementation.
Then, a spatial transformation layer $S$ is employed to warp the
fully-sampled reference $\boldsymbol{x}_{\mathrm{ref}}$ according
to the estimated displacement field $\boldsymbol{\phi}$. Specifically,
in the position $\boldsymbol{p}$, $S\left(\boldsymbol{\phi},\boldsymbol{x}_{\mathrm{ref}}\right)\left[\boldsymbol{p}\right]=\boldsymbol{x}_{\mathrm{ref}}\left[\boldsymbol{p}+\boldsymbol{\phi}\left[\boldsymbol{p}\right]\right]$.
Finally, the spatially aligned image of the fully-sampled reference $\boldsymbol{x}_{\mathrm{ref}}^{\mathrm{A}}=S\left(\boldsymbol{\phi},\boldsymbol{x}_{\mathrm{ref}}\right)$
is fed into the multi-modal reconstruction network together with the
under-sampled target modality $\bar{\boldsymbol{x}}_{\mathrm{tgt}}$,
and the whole multi-modal MRI reconstruction process becomes
\begin{equation}
\hat{\boldsymbol{x}}_{\mathrm{tgt}}=R_{\boldsymbol{\theta}}\left(\bar{\boldsymbol{x}}_{\mathrm{tgt}},S\left(\boldsymbol{\phi},\boldsymbol{x}_{\mathrm{ref}}\right)\right).
\end{equation}
Also, note that usually both fully-sampled reference and target images
are required to estimate their spatial misalignment.
However, in our setting,
the fully-sampled target image is not available before reconstruction,
thus $T_{\boldsymbol{\omega}}$
takes an input of the under-sampled target modality (after zero-filling)
instead of the fully-sampled one.

\subsection{Loss Designs}
\label{section_methods_opt}
To optimize the spatial alignment network
as well as the multi-modal MRI reconstruction network, three loss
functions are used in this work. First, the smoothness loss $\mathcal{L}_{\mathrm{smooth}}$
is imposed in favor of a physically reasonable displacement field.
Then, the reconstruction loss $\mathcal{L}_{\mathrm{rec}}$ asks for
high fidelity of the reconstruction result, which also implicitly
encourages accurate spatial alignment between the target and the reference.
Finally, the registration loss $\mathcal{L}_{\mathrm{reg}}$ is proposed
to provide explicit guidance to optimize the spatial alignment network
and to contribute to high reconstruction quality.

\subsubsection{Smoothness Loss}
It is common to impose smooth prior on the estimated displacement
field $\boldsymbol{\phi}$ in image registration. In our implementation,
the smoothness loss is defined as
$\mathcal{L}_{\mathrm{smooth}}
=\frac{1}{N}\left\Vert \nabla\boldsymbol{\phi}\right\Vert _{2}^{2}$.
For 2D MRI, $\mathcal{L}_{\mathrm{smooth}}$ is expanded as
\begin{multline}
\label{eq_loss_smooth}
\mathcal{L}_{\mathrm{smooth}}=\frac{1}{N}\sum_{\boldsymbol{p}}
\left(\frac{\partial\boldsymbol{\phi}_{x}}{\partial x}\left[\boldsymbol{p}\right]\right)^{2}
+\left(\frac{\partial\boldsymbol{\phi}_{x}}{\partial y}\left[\boldsymbol{p}\right]\right)^{2} \\
+\left(\frac{\partial\boldsymbol{\phi}_{y}}{\partial x}\left[\boldsymbol{p}\right]\right)^{2}
+\left(\frac{\partial\boldsymbol{\phi}_{y}}{\partial y}\left[\boldsymbol{p}\right]\right)^{2}
.
\end{multline}

\subsubsection{Reconstruction Loss}
The image reconstruction loss $\mathcal{L}_{\mathrm{rec}}$ is required
to guide the optimization of the reconstruction network $R_{\boldsymbol{\theta}}$.
It is also possible to optimize the spatial alignment network $T_{\boldsymbol{\omega}}$
with
the hint from the reconstruction loss, as the high-quality reconstruction
of the target usually desires accurate spatial alignment between the two
images. 

In Fig.~\ref{fig_overview}(b), the reconstruction loss $\mathcal{L}_{\mathrm{rec}}$
is applied to the output of the reconstruction network, and its gradient
updates both the reconstruction network $R_{\boldsymbol{\theta}}$
and the spatial alignment network $T_{\boldsymbol{\omega}}$ though
back-propagation.
In particular, we use the popular structural similarity (SSIM)~\cite{wang_multiscale_2003,sriram_end-to-end_2020}
as the reconstruction loss, i.e.,
\begin{equation}\label{eq_loss_rec}
\mathcal{L}_{\mathrm{rec}}= - \mathrm{SSIM}
\left(\hat{\boldsymbol{x}}_{\mathrm{tgt}},
\boldsymbol{x}_{\mathrm{tgt}}\right).
\end{equation}

\subsubsection{Registration Loss}
Using only the reconstruction loss may not be sufficient to optimize
the spatial alignment network, as the back-propagation pathway is
too long for the reconstruction loss $\mathcal{L}_{\mathrm{rec}}$
to reach the spatial alignment network. As the result, we propose
to optimize the spatial alignment network with the direct multi-modal
MRI registration loss $\mathcal{L}_{\mathrm{reg}}$, which aims to
maximize the similarity between the target and the aligned reference images.

\begin{figure*}[!t]
\centerline{\includegraphics[width=0.9\textwidth]{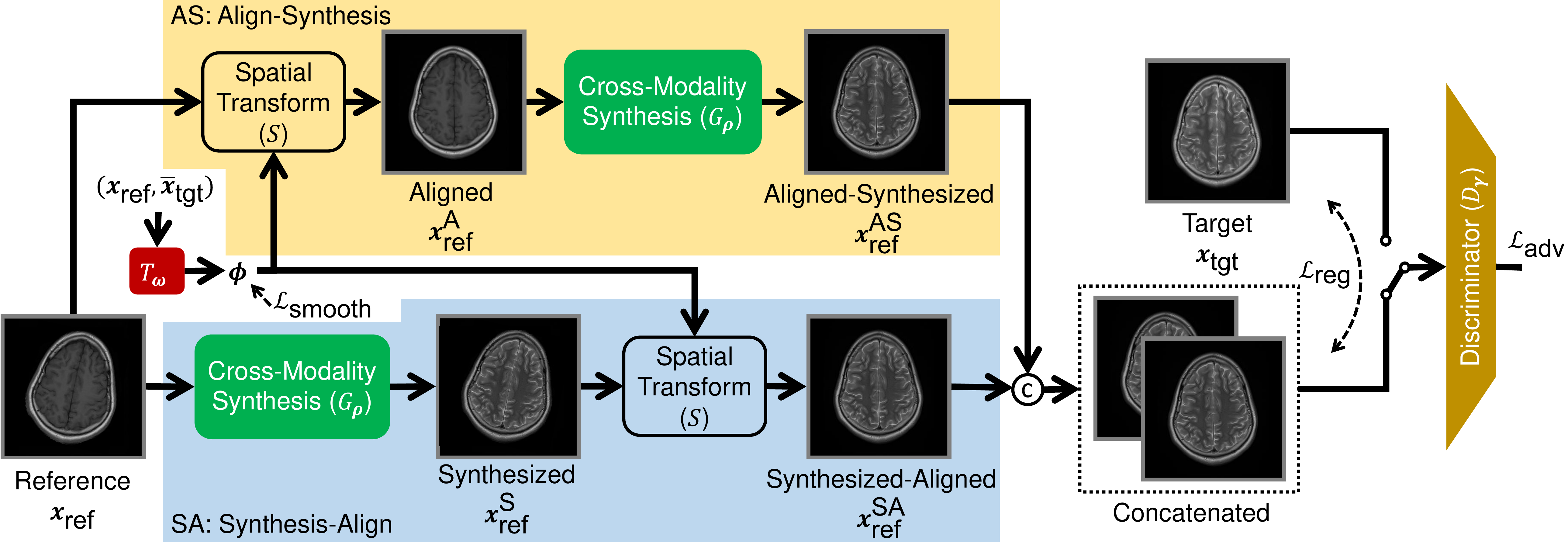}}
\caption{Illustration of the registration loss $\mathcal{L}_{\mathrm{reg}}$.
To measure the cross-modal image similarity between the aligned reference
image and the target modality, a cross-modality synthesis network
is employed to generate the corresponding target modality from the
reference modality. Then we use mono-modality similarity to quantify
the difference between the synthesized and real
target images. We can either place the synthesis network $G_{\boldsymbol{\rho}}$
after and before the spatial transform layer $S$, which derives the
align-synthesis (AS) and synthesis-align (SA) branches, respectively.
Finally, a discriminator is employed in favor of realistic synthesized images.}
\label{fig_Lreg}
\end{figure*}

It is challenging though to design the image similarity metric especially
for multi-modal image registration. Recently, cross-modality image synthesis
has provided a new feasible solution to this problem~\cite{cao_region_2018}.
Instead of directly measuring the similarity
between the target $\boldsymbol{x}_{\mathrm{tgt}}$
and the aligned reference $\boldsymbol{x}_{\mathrm{ref}}^{\mathrm{A}}$,
we first use a cross-modal synthesis network $G_{\boldsymbol{\rho}}$
(marked with green boxes in Fig.~\ref{fig_Lreg})
to produce the synthesized target image
$\boldsymbol{x}_{\mathrm{ref}}^{\mathrm{S}}
=G_{\boldsymbol{\rho}}\left(\boldsymbol{x}_{\mathrm{ref}}\right)$.
Then, the registration tends to align
$\boldsymbol{x}_{\mathrm{ref}}^{\mathrm{S}}$
(instead of $\boldsymbol{x}_{\mathrm{ref}}$) and produces $\boldsymbol{x}_{\mathrm{ref}}^{\mathrm{SA}}=S\left(\boldsymbol{\phi},\boldsymbol{x}_{\mathrm{ref}}^{\mathrm{S}}\right)$.
Next, we can easily calculate $\mathcal{L}_{\mathrm{reg}}^{\mathrm{SA}}=\left\Vert \boldsymbol{x}_{\mathrm{tgt}}-\boldsymbol{x}_{\mathrm{ref}}^{\mathrm{SA}}\right\Vert _{1}$
and treat it as the dissimilarity between $\boldsymbol{x}_{\mathrm{tgt}}$
and $\boldsymbol{x}_{\mathrm{ref}}^{\mathrm{SA}}$. Note that the
superscript $\mathrm{SA}$ indicates the ``synthesis-align (SA)'' strategy.

The cross-modality synthesis network $G_{\boldsymbol{\rho}}$ needs
to be geometry-preserving. That is, before/after synthesis, the same
anatomical structures in the input/output images of $G_{\boldsymbol{\rho}}$
should be spatially aligned, even though the images are of different
modalities in appearance. To this end, we follow
Arar \textit{et	al.}~\cite{arar_unsupervised_2020}
and use the dual-path registration scheme as in Fig.~\ref{fig_Lreg}.
Specifically, in addition to the
previous introduced loss $\mathcal{L}_{\mathrm{reg}}^{\mathrm{SA}}$,
we can place the cross-modality synthesis network after the spatial
transform layer.
In this way, we can derive the ``align-synthesis (AS)'' branch
(in the top of Fig.~\ref{fig_Lreg}),
which is different from the SA branch at the bottom of the figure.
Note that the two branches share
the same cross-modality synthesis network parameters.
In this way, the final registration loss is a mixture of SA and AS:
\begin{equation}\label{eq_loss_reg}
\mathcal{L}_{\mathrm{reg}}=0.5\underbrace{\left\Vert \boldsymbol{x}_{\mathrm{tgt}}-\boldsymbol{x}_{\mathrm{ref}}^{\mathrm{SA}}\right\Vert _{1}}_{\mathrm{SA}}+0.5\underbrace{\left\Vert \boldsymbol{x}_{\mathrm{tgt}}-\boldsymbol{x}_{\mathrm{ref}}^{\mathrm{AS}}\right\Vert _{1}}_{\mathrm{AS}}.
\end{equation}
The above $\mathcal{L}_{\mathrm{reg}}$ also implicitly promotes $\boldsymbol{x}_{\mathrm{ref}}^{\mathrm{SA}}=\boldsymbol{x}_{\mathrm{ref}}^{\mathrm{AS}}$.

\subsection{Implementation Details}
\label{section_methods_details}
In this subsection, we introduce more details of our implementation,
including the adversarial optimization of the cross-modality synthesis network,
and the full algorithm of our multi-modal MRI reconstruction
integrated with the spatial alignment network.

\subsubsection{Adversarial Loss}
To optimize the cross-modality synthesis network $G_{\boldsymbol{\rho}}$,
other than the registration loss $\mathcal{L}_{\mathrm{reg}}$, the
adversarial loss $\mathcal{L}_{\mathrm{adv}}$ is additionally used
to encourage more realistic synthesized images. While the registration
loss $\mathcal{L}_{\mathrm{reg}}$ measures the mean absolute error between
the synthesized and the real target images, the adversarial loss involves
a discriminator $D_{\boldsymbol{\gamma}}$ that forces the synthesized
images to be less distinguishable from the real ones. Following the
adversarial training, the generator $G_{\boldsymbol{\rho}}$ and the
discriminator $D_{\boldsymbol{\gamma}}$ are optimized alternatively.
More specifically, spectral normalization~\cite{miyato_spectral_2018}
and hinge loss are used. Also, the adversarial loss considers both
SA and AS branches as shown in Fig.~\ref{fig_Lreg}:
\begin{equation}\label{eq_loss_adv}
\mathcal{L}_{\mathrm{adv}}=D_{\boldsymbol{\gamma}}\left(\boldsymbol{x}_{\mathrm{tgt}}\right)-0.5D_{\boldsymbol{\gamma}}\left(\boldsymbol{x}_{\mathrm{ref}}^{\mathrm{SA}}\right)-0.5D_{\boldsymbol{\gamma}}\left(\boldsymbol{x}_{\mathrm{ref}}^{\mathrm{AS}}\right).
\end{equation}
The optimization of the cross-modality synthesis network $G_{\boldsymbol{\rho}}$
and discriminator $D_{\boldsymbol{\gamma}}$ can thus be formulated as
\begin{equation}
\boldsymbol{\rho}^{*},\boldsymbol{\gamma}^{*}=\operatorname*{\arg}\operatorname*{\min}_{\boldsymbol{\rho}}\operatorname*{\max}_{\boldsymbol{\gamma}}\mathbb{E}_{p\left(\left(\boldsymbol{x}_{\mathrm{tgt}},\boldsymbol{x}_{\mathrm{ref}}\right)\right)}\alpha\mathcal{L}_{\mathrm{reg}}+\beta\mathcal{L}_{\mathrm{adv}}.
\end{equation}

\subsubsection{Hybrid Supervision over Spatial Alignment Network}
To take advantage of both reconstruction and registration losses
towards better spatial aligning and reconstruction quality,
hybrid supervision is imposed on the spatial alignment network.
The optimizing of the spatial alignment network
solely with the reconstruction loss
may be limited due to the long back-propagation pathway,
while the direct supervision from only the registration loss network may deviate from the final goal of high-quality image reconstruction.
Thus, with both two loss functions, we combine them as the hybrid
supervision upon the spatial alignment network $T_{\boldsymbol{\omega}}$
and the multi-modal reconstruction network $R_{\boldsymbol{\theta}}$:
\begin{equation}
\boldsymbol{\theta}^{*},\boldsymbol{\omega}^{*}=\operatorname*{\arg\min}_{\boldsymbol{\theta},\boldsymbol{\omega}}\mathbb{E}_{p\left(\left(\boldsymbol{x}_{\mathrm{tgt}},\boldsymbol{x}_{\mathrm{ref}}\right)\right)}\mathcal{L}_{\mathrm{rec}}+\lambda\mathcal{L}_{\mathrm{smooth}}+\alpha\mathcal{L}_{\mathrm{reg}}.
\end{equation}

The full training scheme of the multi-modal reconstruction with the spatial alignment
network optimized with hybrid reconstruction and registration losses
is summarized as follows.
\global\long\def\algorithmicrequire{\textbf{Data:}}%
\global\long\def\algorithmicensure{\textbf{Output:}}%
\begin{algorithmic}[1]
\REQUIRE Fully-sampled multi-modal MR images from same studies, $\left\{ \left(\boldsymbol{x}_{\mathrm{tgt}},\boldsymbol{x}_{\mathrm{ref}}\right)\right\} $.
\ENSURE Optimized neural networks, $R_{\boldsymbol{\theta}^{*}}$,
$T_{\boldsymbol{\omega}^{*}}$, $G_{\boldsymbol{\rho}^{*}}$, and
$D_{\boldsymbol{\gamma}^{*}}$. \REPEAT \STATE{$\boldsymbol{x}_{\mathrm{tgt}},\boldsymbol{x}_{\mathrm{ref}}\gets p\left(\left(\boldsymbol{x}_{\mathrm{tgt}},\boldsymbol{x}_{\mathrm{ref}}\right)\right)$}
\STATE{$\bar{\boldsymbol{x}}_{\mathrm{tgt}}\gets\boldsymbol{F}_{\mathrm{0-fill}}^{-1}\boldsymbol{F}_{u}\boldsymbol{x}_{\mathrm{tgt}}$}
\STATE{$\boldsymbol{\phi}\gets T_{\boldsymbol{\omega}}\left(\bar{\boldsymbol{x}}_{\mathrm{tgt}},\boldsymbol{x}_{\mathrm{ref}}\right)$}
\STATE{$\hat{\boldsymbol{x}}_{\mathrm{tgt}}\gets R_{\boldsymbol{\theta}}\left(\bar{\boldsymbol{x}}_{\mathrm{tgt}},S\left(\boldsymbol{\phi},\boldsymbol{x}_{\mathrm{ref}}\right)\right)$}
\STATE{$\boldsymbol{x}_{\mathrm{ref}}^{\mathrm{SA}},\boldsymbol{x}_{\mathrm{ref}}^{\mathrm{AS}}\gets S\left(\boldsymbol{\phi},G_{\boldsymbol{\rho}}\left(\boldsymbol{x}_{\mathrm{ref}}\right)\right),G_{\boldsymbol{\rho}}\left(S\left(\boldsymbol{\phi},\boldsymbol{x}_{\mathrm{ref}}\right)\right)$}
\STATE{Calculate $\mathcal{L}_{\mathrm{smooth}}$, $\mathcal{L}_{\mathrm{rec}}$,
$\mathcal{L}_{\mathrm{reg}}$, and $\mathcal{L}_{\mathrm{adv}}$ with
(\ref{eq_loss_smooth}), (\ref{eq_loss_rec}),
(\ref{eq_loss_reg}), and (\ref{eq_loss_adv}), separately.}
\STATE{$\boldsymbol{\theta}\gets\boldsymbol{\theta}-\eta\partial_{\boldsymbol{\theta}}\mathcal{L}_{\mathrm{rec}}$}
\STATE{$\boldsymbol{\omega}\gets\boldsymbol{\omega}-\eta\partial_{\boldsymbol{\omega}}\left(\mathcal{L}_{\mathrm{rec}}+\lambda\mathcal{L}_{\mathrm{smooth}}+\alpha\mathcal{L}_{\mathrm{reg}}\right)$}
\STATE{$\boldsymbol{\rho}\gets\boldsymbol{\rho}-\eta\partial_{\boldsymbol{\rho}}\left(\alpha\mathcal{L}_{\mathrm{reg}}+\beta\mathcal{L}_{\mathrm{adv}}\right)$}
\STATE{$\boldsymbol{\gamma}\gets\boldsymbol{\gamma}+\eta\partial_{\boldsymbol{\gamma}}\left(\beta\mathcal{L}_{\mathrm{adv}}\right)$}
\UNTIL{convergence}
\end{algorithmic}
where $\lambda$, $\alpha$,
and $\beta$ are the weights for $\mathcal{L}_{\mathrm{smooth}}$,
$\mathcal{L}_{\mathrm{reg}}$, and $\mathcal{L}_{\mathrm{adv}}$,
and $\eta$ is the learning rate.

Hyper-parameters are selected on the validation set
of section~\ref{section_exp_dicom}.
In our implementation, a large weight ($\lambda=1000$) is used
for the smoothness loss $\mathcal{L}_{\mathrm{smooth}}$
of intra-subject registration.
The weight for the synthesis-based image registration loss
$\mathcal{L}_{\mathrm{reg}}$
providing direct guidance to the optimization process
is set to $0.1$ (i.e. $\alpha=0.1$),
as experiments demonstrate that a smaller $\alpha$
is insufficient to align multi-modal MRI,
while a larger $\alpha$ is also harmful
because $\mathcal{L}_{\mathrm{reg}}$
is not exactly consistent with the goal of MRI reconstruction.
Moreover, the weight for our adversarial loss $\mathcal{L}_{\mathrm{adv}}$
is set to $0.01$ (i.e. $\beta=0.01$),
and a smaller $\beta$ may lead to unstable adversarial training.

\section{Experiments}
The experiments are conducted on two MRI datasets.
The first dataset contains real-valued single-coil MRI
stored in digital imaging and communications in medicine (DICOM) format
collected from the large and publicly accessible
NYU fastMRI Initiative database\footnote{\url{https://fastmri.med.nyu.edu/}}%
~\cite{ zbontar_fastmri_2019,knoll_fastmri_2020}
(``DICOM dataset'', Section~\ref{section_exp_dicom}),
and the second one consists of in-house complex-valued multi-coil raw MR images
(``raw dataset'', Section~\ref{section_exp_UII}).
Finally, ablation studies are provided in Section~\ref{section_exp_ablation}.

\begin{figure}[!t]
\centerline{\includegraphics[width=0.8\columnwidth]{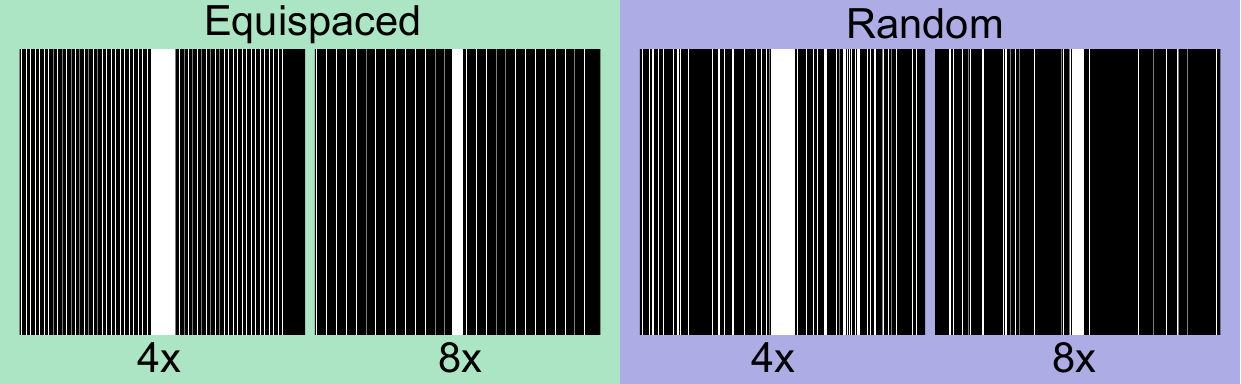}}
\caption{The $k$-space under-sampling patterns.
Numbers at the bottom of individual patterns indicate the acceleration factors.}
\label{fig_masks}
\end{figure}

In both experiments, following the paradigm of fastMRI challenge,
the under-sampled MRI is acquired by down-sampling
corresponding fully-sampled MRI in the $k$-space
with a predefined Cartesian sampling pattern.
More specifically, two typical sampling patterns
--- the random and equispaced sampling patterns --- are used in our experiments,
with a sampling ratio of 25\% or 12.5\%, respectively.
Considering the low-frequency signals contain
most of the energy in the $k$-space, for both random and equispaced patterns,
32\% of the sampling is always allocated to the low frequency,
while the rest sampling is distributed randomly or equispaced.
The under-sampling patterns used in the experiments
are shown in Fig.~\ref{fig_masks}.

The E2E-VarNet~\cite{sriram_end-to-end_2020},
with minimal modification to work with multi-modal MRI,
is employed as the backbone of the reconstruction network.
The U-Net~\cite{ronneberger_u-net_2015} variant similar to that in
Balakrishnan~\textit{et al.}~\cite{balakrishnan_unsupervised_2018}
is used as our spatial alignment network.
Similar to Isola \textit{et al.}~\cite{isola_image_2017},
our generator is the U-Net~\cite{ronneberger_u-net_2015}
empowered with batch normalization~\cite{ioffe_batch_2015},
and our discriminator is a stack of
normalization-activation-convolution paradigms.
Considering all MR images are scanned with large slice spacing and thickness,
2D slices are extracted from 3D volumes for training and evaluation.

Our implementation is mainly based on PyTorch library~\cite{paszke_pytorch_2019}
and the neural networks are trained on Nvidia TITAN RTX graphics cards.
The Adam optimizer~\cite{kingma_adam_2015}
with the learning rate of $\eta=0.0001$
and batch size of 4 is used in all experiments.
Also, the early-stop strategy of $20,000$ iterations,
which take about 3 hours with our hybrid optimization,
is used to prevent over-fitting.
Finally, with the best checkpoint selected on the validation set,
performances on an independent test set are measured and reported.
More implementation details can be found
in our supplementary materials and released code.

\subsection{Experiments on DICOM Dataset}
\label{section_exp_dicom}
We first run our proposed method on the real-valued single-coil MR images
recompiled from the fastMRI database.
While multi-modal MRI is not explicitly provided by fastMRI,
we iterate over all DICOM studies
and select paired T1- and T2-weighted axial brain MRIs.
Finally, 340 pairs of T1- and T2-weighted MR images are extracted.
Among them, 170 pairs of volumes (2720 pairs of slices) are used for training,
68 pairs of volumes (1088 pairs of slices) are taken as the validation set,
and the rest 102 pairs of volumes (1632 pairs of slices) are for testing.
For both T1- and T2-weighted images, the in-plane size is 320$\times$320,
the resolution is $0.68\si{mm}\times0.68\si{mm}$,
and the slice spacing is $5\si{mm}$.
In our experiments,
fully-sampled T1-weighted MRI is used to help the reconstruction of
under-sampled T2-weighted image,
which is consistent with early literature reports%
~\cite{xiang_ultra-fast_2018,kim_improving_2018}.
It is also expected that
T2-weighted MRI can help T1-weighted MRI reconstruction,
and joint reconstruction of both under-sampled T1- and T2-weighted MR images
can achieve high efficiency due to their mutually shared information.

We also find that data augmentation can benefit the reconstruction.
More specifically, for each pair of multi-modal MRI and in each epoch,
a randomly generated deformation field is applied to
both fully-sampled target and reference images.
The deformation field is a combination of
random rotation ($[-0.01\pi,0.01\pi]$),
translation ($[-0.05N,0.05N]$),
and the deformation bicubically interpolated
from $9 \times 9$ control-points
(each assigned with a random displacement in the range of $[-0.02N, 0.02N]$,
where $N$ is the size of the MR images).

Table~\ref{table_dicom_metric} compares the MRI reconstruction performance
with different under-sampling settings and reconstruction methods.
Particularly, ``Single-Modal'' is our single-modal baseline,
where E2E-VarNet~\cite{sriram_end-to-end_2020} reconstructs T2-weighted MRI
without using T1 reference.
``Multi-Modal'' is the multi-modal baseline,
which simply adds T1 reference to E2E-VarNet~\cite{sriram_end-to-end_2020}.
The ``Proposed'' method further integrates the spatial alignment network
with multi-modal E2E-VarNet.
It is observed that the ``Multi-Modal'' baseline
is significantly ($p < 0.01$ with paired \textit{t}-test)
better than ``Single-Modal''
in both peak signal-to-noise ratio (PSNR, the higher, the better)
and SSIM (the higher, the better) in all settings.
Compared to ``Multi-Modal'',
the reconstruction performance can be further improved
with our ``Proposed'' method significantly and consistently.
\begin{table}[!t]
\caption{
Quantitative comparison of reconstruction results on DICOM dataset.
In ``Multi-Modal'' and ``Proposed'' settings, the under-sampled T2-weighted MRI
is reconstructed with helps of fully-sampled T1-weighted image.
For all settings and metrics,
``Multi-Modal'' is significantly better than ``Single-Modal'',
and ``Proposed'' is significantly better than ``Multi-Modal''
($p < 0.01$ with paired \textit{t}-tests).}
\label{table_dicom_metric}
\centering
\begin{tabular}{cc|cc|cc}
\hline\hline
&& \multicolumn{2}{c|}{$4\times$ Acceleration} & \multicolumn{2}{c}{$8\times$ Acceleration} \\
&         & PSNR & SSIM & PSNR & SSIM \\\hline
\multirow{6}{*}{\rotatebox[origin=c]{90}{Equispaced}}
& \multirow{2}{*}{Single-Modal}
& 38.89  &   0.9762   & 37.01     &  0.9674 \\
&& \textpm 1.66 & \textpm 0.0067 & \textpm 1.77 & \textpm 0.0091 \\ \cline{2-6}
& \multirow{2}{*}{Multi-Modal}
& 40.32 & 0.9809 & 38.80 & 0.9755 \\
&& \textpm 1.91 & \textpm 0.0072 & \textpm 2.11 & \textpm 0.0098 \\ \cline{2-6}
& \multirow{2}{*}{Proposed}
& \textbf{40.81} & \textbf{0.9821} & \textbf{39.38} & \textbf{0.9775} \\
&& \textpm 1.92 & \textpm 0.0069 & \textpm 2.06 & \textpm 0.0088 \\\hline
\multirow{6}{*}{\rotatebox[origin=c]{90}{Random}}
& \multirow{2}{*}{Single-Modal}
& 42.72 & 0.9867 & 35.50 & 0.9600 \\
&& \textpm 1.81  & \textpm 0.0043 & \textpm 1.73 & \textpm 0.0103 \\\cline{2-6}
& \multirow{2}{*}{Multi-Modal}
& 43.57 & 0.9882 & 37.42  & 0.9705 \\
&& \textpm 1.95 & \textpm 0.0046 & \textpm 2.00 & \textpm 0.0109 \\\cline{2-6}
& \multirow{2}{*}{Proposed}
& \textbf{43.91} & \textbf{0.9888} & \textbf{38.06} & \textbf{0.9729} \\
&& \textpm 1.94 & \textpm 0.0043 & \textpm 2.04 &\textpm 0.0103 \\\hline\hline
\end{tabular}
\end{table}

\begin{figure}[!t]
\centerline{\includegraphics[width=1\columnwidth]{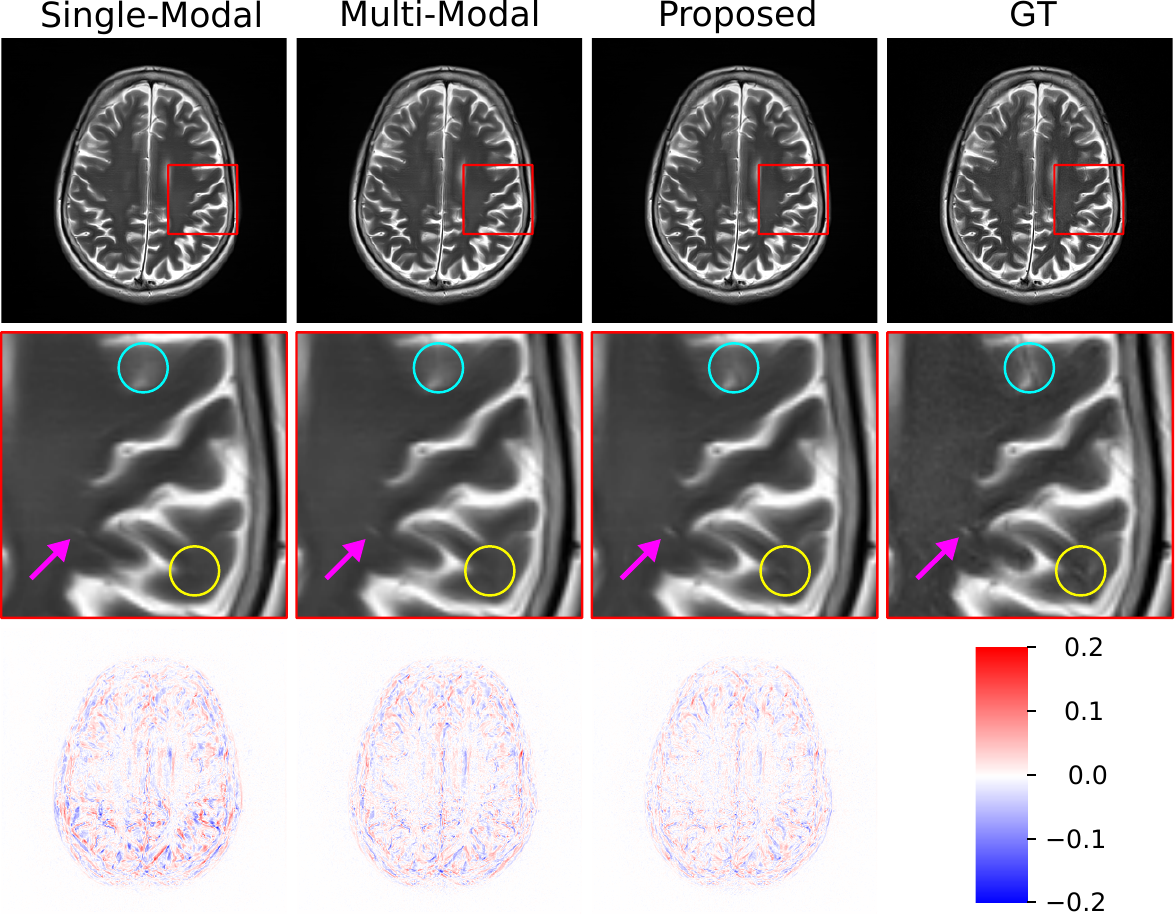}}
\caption{Visual comparison of the reconstructed target MR images
on DICOM dataset.
Results of the equispaced under-sampling pattern
at 8$\times$ acceleration ratio are compared.
The first row is for the reconstructed images,
with the red bounding-boxes zoomed-in in the second row.
The corresponding error maps are provided in the third row.}
\label{fig_dicom_recon}
\end{figure}

Further, a visual comparison of the reconstruction quality
is provided in Fig.~\ref{fig_dicom_recon}.
In general, the multi-modal reconstruction produces clearer MRI images
compared with the single-modal reconstruction,
and our ``Proposed'' method further brings in more details.
For example, the signals in cyan and yellow circles are missing
in ``Single-Modal'' MRI reconstruction.
With a reference modality (``Multi-Modal''),
such signals are less indistinguishable but still vague.
Finally, when the spatial alignment network is integrated,
one can observe a dark vertical line in the cyan circle
and dark scratch-like signals in the yellow circle.
Another example is pointed by the magenta arrows,
where the reconstruction constantly gets clearer from
``Single-Modal'', ``Multi-Modal'' to ``Proposed''.
The error maps,
which subtract the reconstructed target images from the ground-truth,
are also consistent with the quantitative comparisons
in Table~\ref{table_dicom_metric},
as the errors decrease after the spatial alignment network is integrated.

\subsection{Experiments on Multi-Coil Dataset}
\label{section_exp_UII}
In this subsection, we conduct experiments on our raw dataset
which contains in-house complex-valued 24-channel 3T MR images
with raw $k$-space data.
We collect a total of 62 studies, where each study contains paired
T1-weighted fluid-attenuated inversion recovery
(T1-FLAIR, TR=2015ms, TE=12ms)
and T2-weighted
(TR=4226ms, TE=104.8ms) MRI.
The in-plane spacing is $0.7\mathrm{mm} \times 0.7\mathrm{mm}$,
slice thickness is $5 \mathrm{mm}$,
and spatial size of each slice is also 320$\times$320.
Again, 2D slices are extracted from individual volumes,
and T2-weight MR images are supposed to be the target modality.
More specifically, 25 subjects (503 pairs of slices) are used for training,
12 subjects (241 pairs of slices) are for validation,
and the rest 25 subjects (504 pairs of slices) are for testing.
Again, data augmentation is used
considering the limited number of training data.

Quantitative comparison of different reconstruction methods
can be found in Table~\ref{table_uii_metric}.
Compared to the popular ``Single-Modal'' reconstruction,
the information from the reference modality (``Multi-Modal'')
can significantly improve the reconstruction performance
in all settings and metrics.
Also, with our spatial alignment network integrated to
multi-modal reconstruction (``Proposed''),
the metrics of the final reconstruction quality
are further significantly improved.
\begin{table}[!t]
\caption{
Quantitative comparison of the reconstruction results
on private multi-coil MRI dataset with real $k$-space data.
The under-sampled T2-weighted MRI is reconstructed
with helps of the fully-sampled T1-FLAIR image.
For all settings and metrics,
``Multi-Modal'' is significantly better than ``Single-Modal'',
and ``Proposed'' is significantly better than ``Multi-Modal''
($p < 0.01$ with paired \textit{t}-tests).}
\label{table_uii_metric}
\centering
\begin{tabular}{cc|cc|cc}
\hline\hline
&& \multicolumn{2}{c|}{$4\times$ Acceleration} & \multicolumn{2}{c}{$8\times$ Acceleration} \\
&         & PSNR      & SSIM       & PSNR      & SSIM       \\\hline
\multirow{6}{*}{\rotatebox[origin=c]{90}{Equispaced}}
& \multirow{2}{*}{Single-Modal}
& 39.49 & 0.9603 & 35.31 & 0.9309 \\
&& \textpm 1.04 & \textpm 0.0110 & \textpm 0.90 & \textpm 0.0151 \\\cline{2-6}
& \multirow{2}{*}{Multi-Modal}
& 39.60 & 0.9611 & 35.50 & 0.9356 \\
&& \textpm 1.23 & \textpm 0.0114 & \textpm 1.43 & \textpm 0.0190\\\cline{2-6}
& \multirow{2}{*}{Proposed}
&\textbf{40.14} & \textbf{0.9637} & \textbf{36.27} &\textbf{0.9387}\\
&& \textpm 1.24 & \textpm 0.0110 & \textpm 1.43 & \textpm 0.0183 \\\hline
\multirow{6}{*}{\rotatebox[origin=c]{90}{Random}}
& \multirow{2}{*}{Single-Modal}
& 38.59 & 0.9577 & 33.14 & 0.9086 \\
&& \textpm 0.94 & \textpm 0.0107 & \textpm 0.86 & \textpm 0.0161 \\\cline{2-6}
& \multirow{2}{*}{Multi-Modal}
& 38.90 & 0.9580 & 34.11 & 0.9240 \\
&& \textpm 1.31 & \textpm 0.0119 & \textpm 1.54 & \textpm 0.0231 \\\cline{2-6}
& \multirow{2}{*}{Proposed}
&\textbf{39.45} & \textbf{0.9603} & \textbf{34.65} & \textbf{0.9291} \\
&& \textpm 1.24 & \textpm 0.0111 & \textpm 1.52 & \textpm 0.0222 \\\hline\hline
\end{tabular}
\end{table}

\begin{figure}[!t]
\centerline{\includegraphics[width=1\columnwidth]{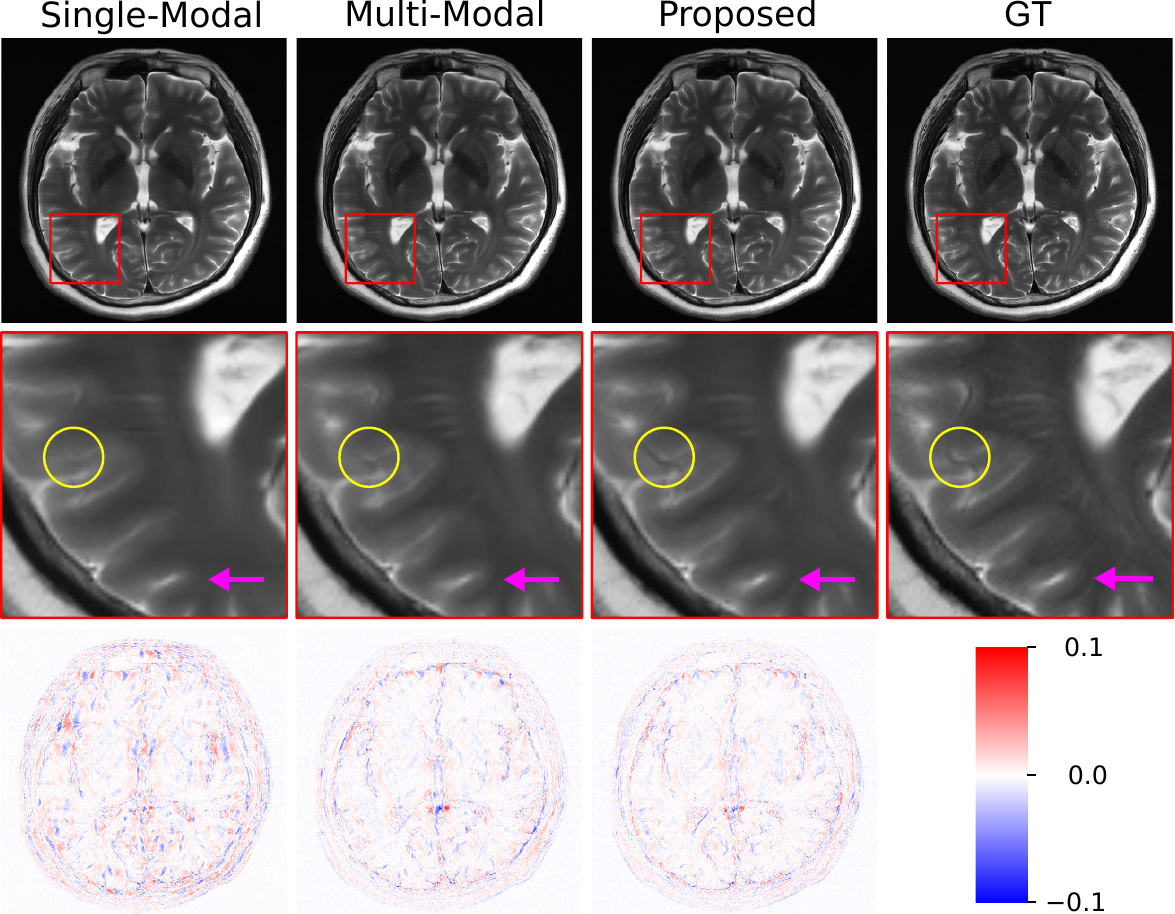}}
\caption{
Visual comparison of the reconstructed images,
zoomed-in views, and error maps on our raw dataset.
T1-FLAIR MR images are used to help the reconstruction of T2-weighted MRI
with random under-sampling pattern
at 8$\times$ acceleration ratio.}
\label{fig_uii_recon}
\end{figure}

Fig.~\ref{fig_uii_recon} visually illustrates the reconstruction results
and corresponding error maps of T2-weighted MRI under-sampled
with the random sampling pattern at 8$\times$ acceleration ratio.
Generally speaking, visual quality and error maps monotonically get better
from ``Single-Modal'', ``Multi-Modal'' to ``Proposed'',
which is consistent with Table~\ref{table_uii_metric}.
More specifically, comparing  ``Multi-Modal'' and ``Single-Modal'',
the tissue boundaries are sharper in MR images
reconstructed with the reference modality,
including but not limited to the white matter (WM) / gray matter (GM) boundary
pointed by magenta arrows.
Comparing ``Proposed'' and ``Multi-Modal'',
our proposed method brings in more details,
like the S-shaped dark signal in yellow circles.

\subsection{Ablation Study}
\label{section_exp_ablation}
In this section,
we analyze the effects of our hybrid supervision over spatial alignment network
(Section \ref{section_exp_ablation_loss}),
the impact of spatial misalignment upon the image quality
(Section \ref{section_exp_ablation_misalignment}),
and computational complexity
(Section \ref{section_exp_ablation_complexity}).

\subsubsection{Hybrid supervision}
\label{section_exp_ablation_loss}
\begin{figure}[!t]
\centerline{\includegraphics[width=1\columnwidth]{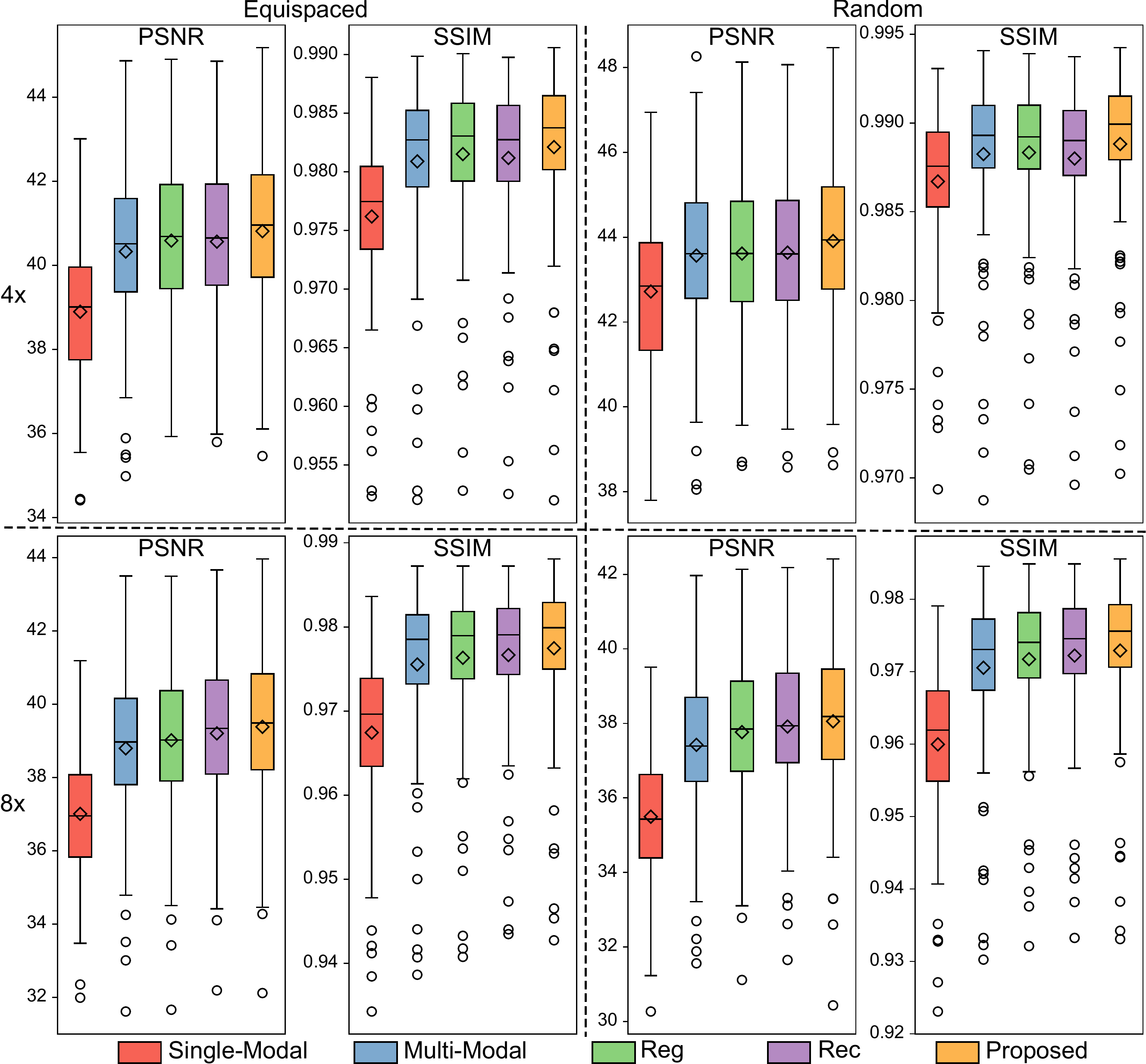}}
\caption{
Quantitative comparison of MRI reconstruction (in PSNR/SSIM)
with different methods and loss functions on the DICOM dataset.
In each box plot, the average value is marked with the diamond symbol.}
\label{fig_dicom_box}
\end{figure}

\begin{figure}[!t]
\centerline{\includegraphics[width=\columnwidth]{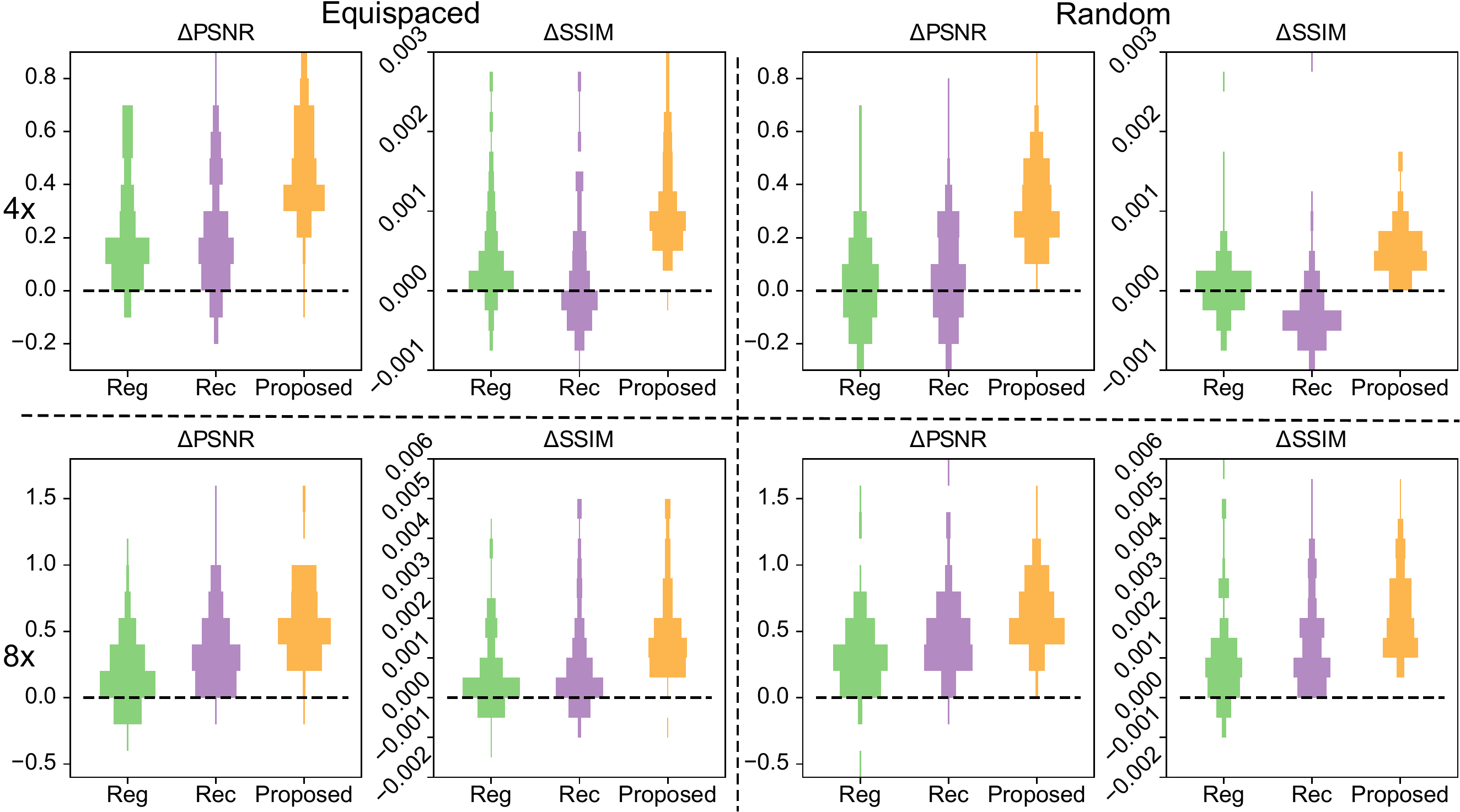}}
\caption{Performance improvement (in PSNR/SSIM)
from the spatial alignment network with different optimization strategies
over multi-modal reconstruction without spatial alignment on the DICOM dataset.}
\label{fig_dicom_hist}
\end{figure}

\begin{figure}[!t]
\centerline{\includegraphics[width=\columnwidth]{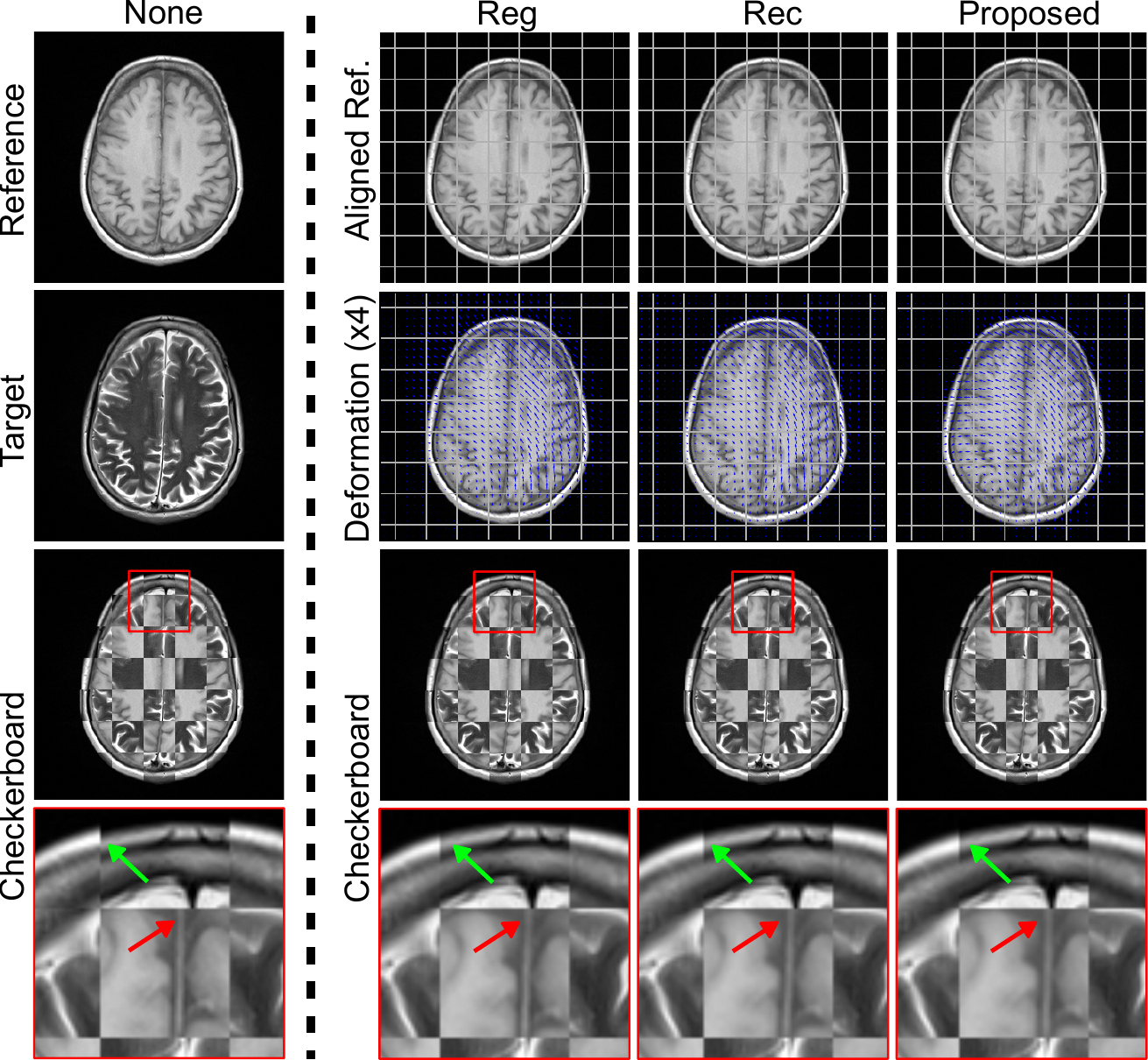}}
\caption{
Effects of the spatial alignment network.
The first row shows the original reference images (first column)
and the warped ones with different optimization strategies (last three columns).
In the second row, the first column is the fully-sampled target,
and the rest columns are the deformation fields.
The third row contains checkerboard visualizations of
the spatial misalignment between the target
and the original/warped reference images,
the zoomed-in views of which are presented in the last row.
The results of the random under-sampling pattern
at 8$\times$ acceleration ratio are compared.
Note the displacement fields are amplified by four times
for better visualization.
}
\label{fig_dicom_align}
\end{figure}

Our hybrid supervision over the spatial alignment network,
which consists of a common reconstruction loss $\mathcal{L}_{\mathrm{rec}}$
and a direct cross-modality-synthesis-based registration loss
$\mathcal{L}_{\mathrm{reg}}$,
is of vital importance to the network optimization.
To prove the necessity and superiority of such hybrid supervision,
we perform experiments
with the spatial alignment network $T_{\boldsymbol{\omega}}$
and the reconstruction network $R_{\boldsymbol{\theta}}$ optimized
with solely the reconstruction loss $\mathcal{L}_{\mathrm{rec}}$
(``Rec'' in Fig.~\ref{fig_dicom_box},
\ref{fig_dicom_hist}, and \ref{fig_dicom_align})
or the registration loss $\mathcal{L}_{\mathrm{reg}}$
(``Reg'' in Fig.~\ref{fig_dicom_box},
\ref{fig_dicom_hist}, and \ref{fig_dicom_align}).
For training with the reconstruction loss $\mathcal{L}_{\mathrm{rec}}$ only
(``Rec''),
we have disabled the registration loss $\mathcal{L}_{\mathrm{reg}}$
and the adversarial training $\mathcal{L}_{\mathrm{adv}}$.
For training with the reconstruction loss $\mathcal{L}_{\mathrm{reg}}$ only
(``Reg''),
first our spatial alignment network $T_{\boldsymbol{\omega}}$
is optimized without $\mathcal{L}_{\mathrm{rec}}$,
and then we freeze $T_{\boldsymbol{\omega}}$
to train the reconstruction network $R_{\boldsymbol{\theta}}$
with $\mathcal{L}_{\mathrm{rec}}$.

Fig.~\ref{fig_dicom_box} quantitatively demonstrates
the superiority of our hybrid supervision
and Fig.~\ref{fig_dicom_hist} supports its robustness on the DICOM dataset.
In Fig.~\ref{fig_dicom_box},
the performances of different methods, settings, and metrics
are presented in box plots,
with black bars and diamonds showing the median and mean values.
The box plots from left to right are for single-modal reconstruction (``Single-Modal'', pink), multi-modal reconstruction without spatial alignment network (``Multi-Modal'', blue), and multi-modal reconstruction with spatial alignment network trained with solely $\mathcal{L}_{\mathrm{reg}}$ (``Reg'', green), solely $\mathcal{L}_{\mathrm{rec}}$ (``Rec'', purple), or our proposed hybrid supervision (``Proposed'', yellow).
Comparing single-modal reconstruction (``Single-Modal'') with multi-modal reconstruction (``Multi-Modal'', ``Rec'', ``Reg'', and ``Proposed''), the latter ones are always better than the MR images reconstructed without a reference modality.
Among multi-modal reconstruction methods,
the spatial alignment network
trained with solely $\mathcal{L}_{\mathrm{reg}}$ (``Reg'')
or $\mathcal{L}_{\mathrm{rec}}$ (``Rec'')
not necessarily benefits multi-modal reconstruction (``Multi-Modal'')
(e.g. SSIM of $4\times$ random under-sampling pattern),
while our hybrid supervision (``Proposed'') can always
perform better than all other methods.
Fig.~\ref{fig_dicom_hist} plots the distribution of
the subject-wise performance gains brought by the spatial alignment network,
which are optimized with different loss functions
(``Rec'', ``Reg'', and ``Proposed'')
and compared to multi-modal reconstruction
without the spatial alignment network (``Multi-Modal'').
It reveals that the optimization with
solely $\mathcal{L}_{\mathrm{reg}}$ (``Reg'')
or $\mathcal{L}_{\mathrm{rec}}$ (``Rec'')
is suboptimal and somehow unstable
where the portion of the subjects with positive performance gain
varies from 20\% to 100\%.
For example,
with $4\times$ acceleration, random under-sampling pattern, and SSIM metric,
the performance improvement brought by spatial alignment network
trained with the ``Rec'' strategy is negative for most of the subjects,
which is absolutely not acceptable.
On the contrary, the spatial alignment network
trained with hybrid supervision (``Proposed'')
not only performs better than other optimization strategies
(``Reg'' and ``Reg'')
but also keeps very robust where over 98\% subjects can benefit.

It is also necessary to explore the displacement fields
generated by the spatial alignment network.
Fig.~\ref{fig_dicom_align} visualizes the estimated displacement fields
and the aligned reference images.
The displacements are amplified for a factor of 4
and rendered in blue arrows for a clear illustration.
The two rows at the bottom are the fully-sampled target images
and the ground-truth reference modality in checkerboard visualization.
In the zoomed-in views (last row),
the spatial misalignment between the two modalities is clearly visible,
in that the boundary of the skull is mismatched (pointed by green arrows)
and the discontinuity also exists in the brain center-line
(pointed by red arrows).
In both ``Reg'' and ``Proposed'' particularly,
the reference image can be well aligned with the target modality,
while ``Rec'' is less accurate in aligning brain center-lines (pointed by red arrows as in the last row of the figure).
The comparisons underscore our argument that the long-way back-propagated from the reconstruction loss can be insufficient to optimize the spatial alignment network.
We also note that, given the example in Fig.~\ref{fig_dicom_align},
the spatial misalignment appears to be
subtle rotation in the anti-clockwise direction,
which can be attributed to the unexpected motion of the subject.

\subsubsection{Degree of Spatial Misalignment}
\label{section_exp_ablation_misalignment}
\begin{figure}[!t]
\centerline{\includegraphics[width=\columnwidth]{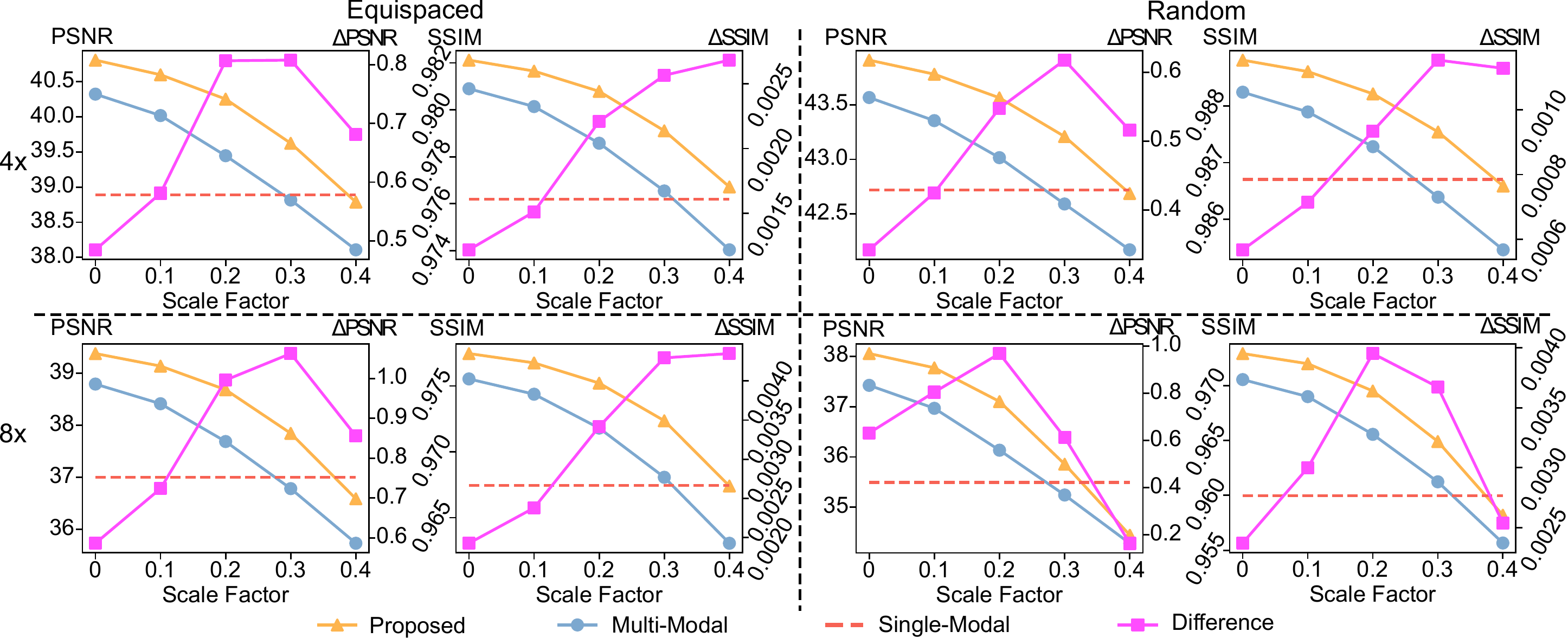}}
\caption{
Quantitative comparison of multi-modal MRI reconstruction
with varying degrees of simulated spatial misalignment.
For all subplots, left y-axes are for reconstruction performances
(``Proposed'', ``Multi-Modal'', and ``Single-Modal'')
while the right y-axes are for the ``Difference''
between ``Proposed'' and ``Multi-Modal''.
}
\label{fig_dicom_diff}
\end{figure}

Multi-modal reconstruction assisted with the spatial alignment network
is more robust to different degrees of spatial misalignment.
To prove this, we simulate different degrees of spatial misalignment
to our DICOM dataset
and then evaluate the reconstruction methods
optimized in Section~\ref{section_exp_dicom}.
The simulated spatial misalignment
is similar to data augmentation,
including random rotation ($[-0.01\pi\sigma,0.01\pi\sigma]$),
translation ($[-0.05N\sigma,0.05N\sigma]$),
and deformation field bicubically interpolated from
$9 \times 9$ control-points (with displacements uniformly sampled
within $[-0.02N\sigma,0.02N\sigma]$ in both directions.
Here $N$ is the size of the MR images
and $\sigma$ is the factor
controlling the degree of spatial misalignment).
Also, note that the simulated spatial misalignment
applies to the reference images only,
while data augmentation warps both reference and target images.
Fig.~\ref{fig_dicom_diff}
compares the reconstruction performances (left y-axes)
and performance gains of the proposed method (right y-axes)
with varying degrees of simulated spatial misalignment (x-axes).
The ``Proposed'' method is plotted in orange triangles,
the ``Multi-Modal'' reconstruction without the spatial alignment
is plotted in blue circles,
the magenta rectangles are for the performance improvement
brought by the spatial alignment network (``Difference''),
and the pink dashed lines are for the ``Single-Modal'' reconstruction.
The curves in Fig.~\ref{fig_dicom_diff}
clearly indicate that the performance of both ``Proposed'' and ``Multi-Modal''
monotonically drops with larger spatial misalignment.
It is also interesting to see that, with spatial misalignment increasing,
the ``Difference'',
or the performance improvement between ``Proposed'' and ``Multi-Modal'',
first increase and then decreases.

\subsubsection{Computational Complexity Comparison}
\label{section_exp_ablation_complexity}
Other than accuracy, computational complexity is also important.
Table~\ref{table_complexity} lists four important measures of the complexity, including the number of parameters, the amount of multiply–accumulates (MACs),
the inference time and memory requirement of $G_{\boldsymbol{\rho}}$, $D_{\boldsymbol{\gamma}}$, $T_{\boldsymbol{\omega}}$ and $R_{\boldsymbol{\theta}}$.
More specifically, MACs are measured with the ptflops%
\footnote{%
\url{https://github.com/sovrasov/flops-counter.pytorch}} tool,
while the inference time and the memory footprint
are measured on an Nvidia TITAN RTX GPU.
For all networks,
the computational complexity with both DICOM and raw datasets is measured.
From Table~\ref{table_complexity},
it is obvious that our cross-modality synthesis network $G_{\boldsymbol{\rho}}$
and the reconstruction network $R_{\boldsymbol{\theta}}$
cost much more than the other two networks.
Moreover, considering that $G_{\boldsymbol{\rho}}$
is only used in the offline training,
our spatial alignment network $T_{\boldsymbol{\omega}}$,
which is an ultra-light module,
is the only overhead compared with the multi-modal MRI reconstruction method
without spatially aligning the reference image.

\begin{table}[!t]
\caption{
Computational complexity of neural networks used in our proposed method.}
\label{table_complexity}
\centering
\begin{tabular}{c|c|c|c|c|c}
\hline\hline
Coils \& Size & Complexity &
$G_{\boldsymbol{\rho}}$ & $D_{\boldsymbol{\gamma}}$ &
$T_{\boldsymbol{\omega}}$ & $R_{\boldsymbol{\theta}}$ \\\hline
\multirow{2}{*}{1-Coil MRI}
& MACs (G)       & 88.33 & 17.63 & 12.80 & 55.95 \\\cline{2-6}
& Parameters (M) & 22.88 &  3.51 &  0.72 & 20.12 \\\cline{2-6}
\multirow{2}{*}{$320\times320$}
& Time (ms)      & 16.01 & 3.06  & 4.27  & 31.61  \\\cline{2-6}
& Memory (MiB)   & 179.63 & 52.39 & 78.06 & 42.13 \\\hline
\multirow{2}{*}{24-Coil MRI}
& MACs (G)       & 88.33 & 17.63 & 14.15 & 87.29 \\\cline{2-6}
& Parameters (M) & 22.88 &  3.51 &  0.73 & 20.12 \\\cline{2-6}
\multirow{2}{*}{$320\times320$}
& Time (ms)      & 15.99 & 3.06  & 4.49  & 69.66  \\\cline{2-6}
& Memory (MiB)   & 179.74 & 52.39 & 96.03 & 112.89 \\\hline\hline
\end{tabular}
\end{table}

\section{Discussion and Conclusion}
In this manuscript, to deal with the subtle spatial misalignment between
MRI sequences, which is unfortunately prevalent even in the same study, we
propose to explicitly align the reference modality with the target
modality, and feed them together to the deep network for the reconstruction of the target.
More specifically, the spatial alignment network takes fully-sampled
reference MRI and under-sampled target image as the input and outputs
the displacement field to align the reference modality. Also, a hybrid
optimization strategy is proposed to provide strong training
supervision for the spatial alignment network.
Experiments on both the DICOM and raw MRI datasets demonstrate that the spatial
alignment network not only improves overall reconstruction performance
but also is robust to bring performance gain to almost all subjects.
The proposed spatial alignment network is flexible,
and theoretically,
it can be plug-in into any multi-modal reconstruction network.

We believe that the spatial alignment network is an important
yet missing component towards practical multi-modal reconstruction.
Further research directions include extending our solution
to more organs and 3D MR images,
while in this manuscript we only work on 2D cases due to large slice spacing.
Also, it is a promising way
to alternatively align different sequences and reconstruct MRI,
which is suitable to work with state-of-the-art unrolled reconstruction methods.
Finally, as a general plug-and-play module,
the spatial alignment network has great potential in improving the performance
of joint reconstruction with multiple ($\geq3$) MRI sequences
or joint k-space under-sampling pattern learning.

\bibliographystyle{IEEEtran}
\bibliography{mybibfile}
\end{document}